\documentclass[prb,twocolumn,aps,showpacs,fixfloats]{revtex4}
\usepackage{graphicx}
\usepackage{bm}
\usepackage{amsmath,amssymb}
\usepackage{subfigure}
\usepackage{float}
\usepackage{latexsym}
\usepackage{epstopdf}
\usepackage{color}
\usepackage{enumerate}
\usepackage{pdfpages}

\newcommand{\s}{\scriptscriptstyle}

\begin{document}

\title {Effect of extended confinement on the structure of edge channels in the quantum  anomalous Hall effect  }

\author{Z. Yue and M. E. Raikh }

 \affiliation{Department of Physics and
Astronomy, University of Utah, Salt Lake City, UT 84112, USA
}

\begin{abstract}
Quantum anomalous Hall (QAH) effect  in the films with nontrivial band structure
accompanies the ferromagnetic transition in the system of magnetic dopants. Experimentally, 
the QAH transition manifests itself as a jump in the dependence of longitudinal resistivity on a weak external magnetic
field. Microscopically, this jump originates from the emergence of a chiral edge mode on one side 
of the ferromagnetic transition.
We study analytically the  effect of an extended  confinement
on the structure of the edge modes.  We employ the simplest model of
the extended confinement in the form of potential step next to
the hard wall. It is shown that, unlike the conventional quantum Hall effect, 
where all edge channels are chiral,
in QAH effect, a complex structure of the boundary leads to nonchiral edge modes  which are present
on  both sides of the ferromagnetic transition.
Wave functions of nonchiral modes are different above and below the transition:
on the ``topological" side, where the chiral edge mode is supported, nonchiral modes are ``repelled" from the boundary, i.e.
they are much less localized than on the ``trivial" side. Thus, the disorder-induced scattering into these
modes will boost the extension of the chiral edge mode.
The prime experimental manifestation of nonchiral modes is
that, by  contributing to longitudinal resistance, they smear
the QAH transition.
\end{abstract}
\maketitle

\begin{figure}
\includegraphics[width=84mm]{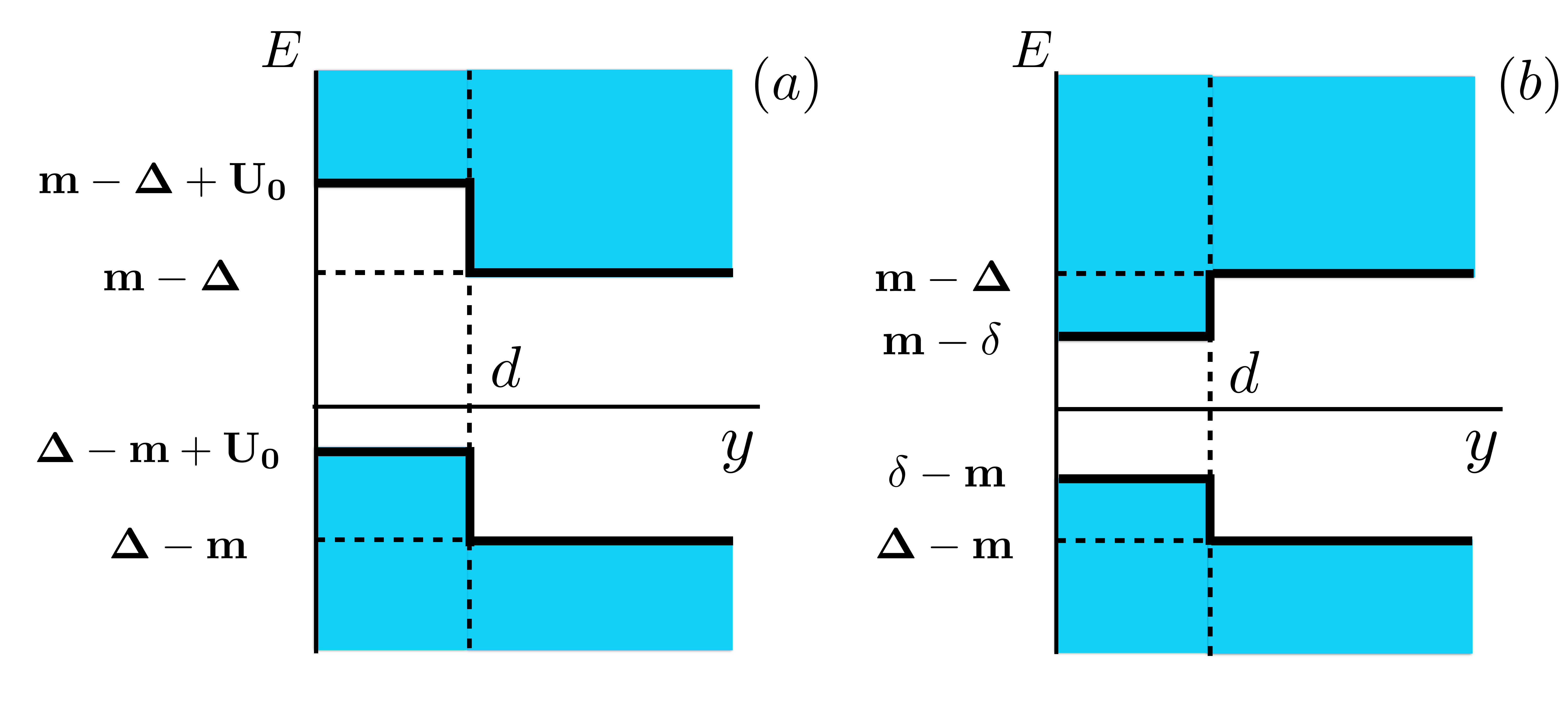}
\caption{ Two variants of the extended confinement: (a) potential step next to the hard wall, and (b) step in the gapwidth next to the hard wall.}
\label{figure1}
\end{figure}

\section{Introduction}
Quantum anomalous Hall effect
is achieved by doping the films possessing nontrivial band structure
with magnetic impurities.\cite{1,2,3,Pioneering,Chekelsky1,UCLA1,Robust,PennState,Tokura1,UCLA2,Goldhaber,Moodera1,Moodera2,Molenkamp,Zeldov}
This doping gives rise to a spontaneous magnetization caused by exchange between the impurities.
The most exciting consequence of this magnetization is that
the associated spin splitting
results
in the band inversion.
Magnetization-induced band inversion was predicted theoretically in
Refs. \onlinecite{YSWu2006},~\onlinecite{2008}.
First experiments\cite{1,2,3} indicated that there is a jump in non-diagonal component, $\sigma_{xy}$, of the conductivity
at ferromagnetic transition  confirming the theoretical prediction.
Very recently\cite{Goldhaber,Moodera1}, upon  improving the quality of the samples, a very accurate quantization
of  $\sigma_{xy}$ was demonstrated.

In experiments\cite{1,2,3,Pioneering,Chekelsky1,UCLA1,Robust,PennState,Tokura1,UCLA2,Goldhaber,Moodera1,Moodera2,Molenkamp,Zeldov},
the ferromagnetism is switched on and off by application of a weak external field. The observed quantized steps
in non-diagonal resistance look similar to the steps observed in conventional quantum Hall effect only in much weaker
external fields.
One of the conclusions which can be drawn from these experimental studies is that the
structure of the edge states plays a crucial role in
achieving an almost zero longitudinal resistance, $\rho_{xx}$.

On the theoretical side, it was demonstrated numerically in Ref. \onlinecite{EdgeTransportQAH} that the dispersion law of the
edge states in realistic multilayer QAH structure contains  nonchiral edge modes along with
a chiral one. It was also demonstrated in Ref. \onlinecite{EdgeTransportQAH}
that coexistence of chiral and nonchiral edge modes
leads to a finite longitudinal resistance.
In order to  suppress the contribution
of nonchiral channels to $\rho_{xx}$,  in experiment Ref. \onlinecite{Goldhaber} it was proposed to localize them by disorder. Indeed, for nonchiral edge modes, the backscattering and, consequently, the interference is allowed. This interference, on the other hand,  is the origin of the quantum localization.

In theory, the question whether or not a given band structure allows a chiral
edge state is decided by calculating  the Chern number. Naturally, this calculation does not
answer a question whether or not this band structure supports nonchiral  in-gap edge modes.
Alternative microscopic approach \cite{Sonin2010,Fabian2012,Platero2013,Durnev}
to the issue of edge states confirms the prediction about their presence or absence made on the basis of Chern number calculation. This microscopic approach also allows to calculate analytically
the modification of the wave function of the chiral edge state due to the orbital action of magnetic field and, even, to trace how this edge state transforms into the quantum Hall edge state upon increasing the field.
However, microscopic approach \cite{Sonin2010,Fabian2012,Platero2013,Durnev}
equally does not reproduce the nonchiral modes within the envelop-function description.

The Hamiltonian describing the
gapped edge spectrum in QAH has a $2\times 2$
matrix form.\cite{YSWu2006} This is in contrast
to the conventional spin-orbit $4\times 4$ Hamiltonian\cite{Science} describing
the states in HgTe-based quantum wells. The reason is that the transition
between inverted and trivial band structures due to magnetization takes
place only for one spin projection.
 As a consequence of the matrix form of the Hamiltonian, the in-gap eigenstates are characterized
 by two decay lengths. Edge state is allowed if the two corresponding eigenvectors can be combined to satisfy the hard-wall boundary condition.\cite{Dyakonov} It appears that only ``nontrivial" band structure allows such combination.

In the present paper we demonstrate that nonchiral edge modes emerge
naturally upon generalization of the microscopic approach \cite{Sonin2010,Fabian2012,Platero2013,Durnev} to the case
of the extended confinement. In fact, we employ the simplest model
of the extended confinement in the form of a step next to the hard wall.
We demonstrate that both chiral and nonchiral modes emerge as
solutions of the same characteristic equation. The wave functions
of nonchiral modes oscillate within the step before decaying into
the bulk.
Within the simplest model considered, we compare, for the same confinement,
nonchiral edge modes for inverted band structure, supporting the chiral mode,
and for ``trivial" band structure. Our main finding is that, for  ``trivial" band structure,
the nonchiral modes have a lower threshold. Nonchiral modes with inverted
band structure are more extended.   Disorder-induced scattering into these
states extends the localization length of the chiral edge mode.

\section{Edge modes in the presence of a step}

\subsection{Hard wall}
\begin{figure}
\includegraphics[width=58mm]{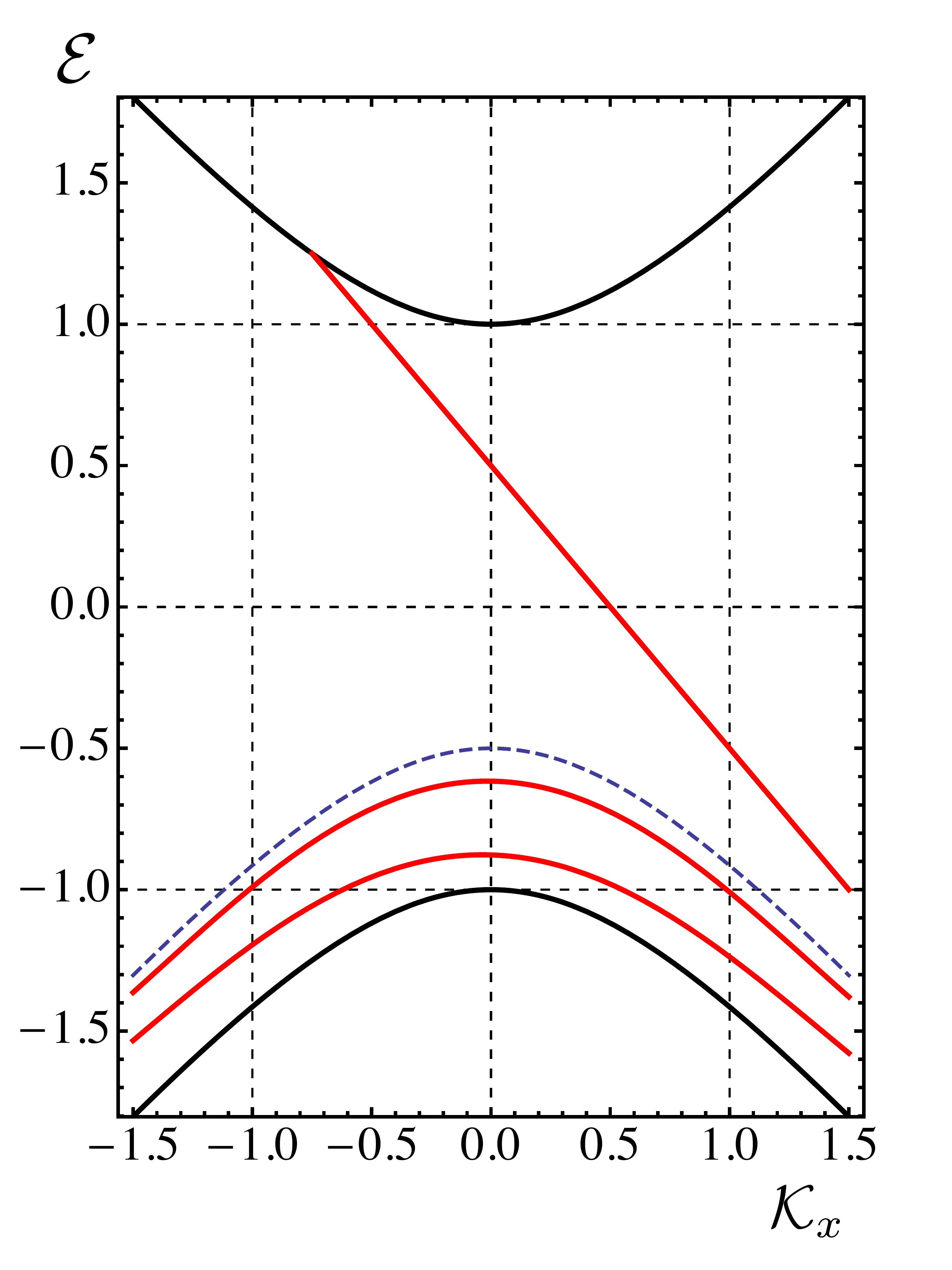}
\caption{(Color online)  The dispersions of the modes propagating along the boundary $y=0$ is plotted from Eqs. (\ref{transcendental}) and (\ref{transcendental2}) for dimensionless step height $\tilde{U}_0=0.5$ and dimensionless step width $\tilde{d}=6$. The spectrum of the edge mode and of two nonchiral modes is shown with red lines. Bulk  spectrum (black) and the spectrum in the step region (dashed) are also shown.}
\label{figure2}
\end{figure}

To introduce notations, we briefly review the structure of the bulk QAH Hamiltonian\cite{2008}.
It emerges from the conventional $4 \times 4$ Hamiltonian\cite{Science}
\begin{equation}
\hat{H}_{\text{eff}}=\begin{pmatrix} \hat{h}({\bf k}) & 0 \\ 0 & \hat{h}^{\ast}(-{\bf k})\end{pmatrix}
\end{equation}
where $\hat{h}({\bf k})$ is a $2\times 2$ matrix defined as
$\hat{h}({\bf k})=A(k_x \sigma_x+k_y \sigma_y)+(m+Bk^2)\sigma_z$,  while $\sigma_y$, $\sigma_z$ are the Pauli matrices
acting in the pseudospin (electron-heavy hole) subspace.
Upon adding the exchange
\begin{equation}
\hat{h}_{\text{exch}}=\begin{pmatrix} \Delta & 0 & 0 & 0 \\ 0 &-\Delta & 0 & 0
 \\ 0 & 0 & -\Delta & 0 \\ 0 & 0 & 0 & \Delta
\end{pmatrix}
\end{equation}
the two blocks become inequivalent
\begin{align}
\label{3}
\hat{h}({\bf k}) &\rightarrow \begin{pmatrix} m+\Delta+Bk^2 & -A(k_x-i k_y) \\ - A(k_x+i k_y)& -m-\Delta-Bk^2\end{pmatrix}  \\
\label{4}
\hat{h}^{\ast}(-{\bf k}) & \rightarrow  \begin{pmatrix} m-\Delta+Bk^2 & A(k_x+i k_y) \\  A(k_x-i k_y) & -m+\Delta-Bk^2\end{pmatrix}
\end{align}
Near $m=\Delta$ the band inversion takes place only in the second block.
Thus the transition can be swept through by applying a weak magnetic field,
since the field controls the parameter, $\Delta$.

It follows from Eq. (\ref{4}) that at the transition $m=\Delta$ the Hamiltonian possesses only
a single spatial scale,
\begin{equation}
l_0=\frac{B}{A}.
\end{equation}
 Away from the transition, a new spatial scale,
 \begin{equation}
l_{\Delta}=\frac{A}{m-\Delta},
\end{equation}
 appears.
The theory is greatly simplified by the fact that the first scale is much smaller than the second one.
In terms of the edge states, for a given, say positive, sign of $B$, the edge state is present for
$m<\Delta$ and is absent for $m>\Delta$.
To see this, consider the two  eigenvectors of $\hat{h}^{\ast}(-{\bf k})$
propagating, as $\exp (ik_xx)$, along the boundary $y=0$  and decaying, as $\exp (-q y)$, into the bulk, $y>0$.
For these eigenvectors, the elements of corresponding pseudospinors are related as
\begin{align}
\label{I}
&[m-\Delta+B(k_x^2-q^2)-E]\alpha+A(k_x-q)\beta=0, \nonumber\\
&[m-\Delta+B(k_x^2-q^2)+E]\beta-A(k_x+q)\alpha=0.
\end{align}
With $l_0 \ll |l_{\Delta}|$, the $q$-values for the two eigenvectors differ strongly, and the
expressions for them have a simple form
\begin{align}
\label{II}
q_0=\frac{1}{l_0},~~~q_{\s \Delta}=\frac{1}{|l_{\Delta}|}\sqrt{1+(l_\Delta k_x)^2-\Bigg(\frac{E l_\Delta}{A}\Bigg)^2}.
\end{align}
Note that, by virtue  of the condition $l_0 \ll |l_{\Delta}|$, the nonparabolicity parameter $B$
does not enter into $q_{\s \Delta}$.
Substituting Eq. (\ref{II}) into  Eq. (\ref{I}), we find the form of the corresponding eigenvectors
\begin{align}
\label{III}
\Psi_0&=\begin{pmatrix} -1 \\ 1\end{pmatrix} \exp\Big[i k_x x-q_0 y\Big],\\
\label{III'}
\Psi_\Delta &=\begin{pmatrix} 1 \\ \frac{A(k_x +q_{\s \Delta})}{m-\Delta+E}\end{pmatrix} \exp\Big[i k_x x-q_{\s \Delta} y\Big].
\end{align}
To satisfy the hard-wall boundary condition,
both components of the linear combination of the eigenvectors Eqs. (\ref{III}), (\ref{III'})
should turn to zero at $y=0$.  This amounts to the requirement
\begin{equation}
\label{IV}
1+\frac{A(k_x +q_{\s \Delta})}{m-\Delta+E}=\frac{2(Ak_x+E)}{m-\Delta+E+A(k_x-q_{\s \Delta})}=0.
\end{equation}
One immediately concludes from Eq. (\ref{IV}) that the dispersion law of the chiral edge mode is linear
\begin{equation}
\label{chiral}
 E=-Ak_x.
\end{equation}
However, this conclusion applies only on one side of the transition, namely, for $(m-\Delta)<0$. Indeed, as it follows from Eq. (\ref{II}), for $E=-Ak_x$, we have $q_{\s \Delta}=A/|m-\Delta|$.
Therefore, for positive $m-\Delta$, the denominator in Eq. (\ref{IV}) turns to zero together with the numerator, so that the boundary
condition cannot be satisfied.

\begin{figure}
\includegraphics[width=84mm]{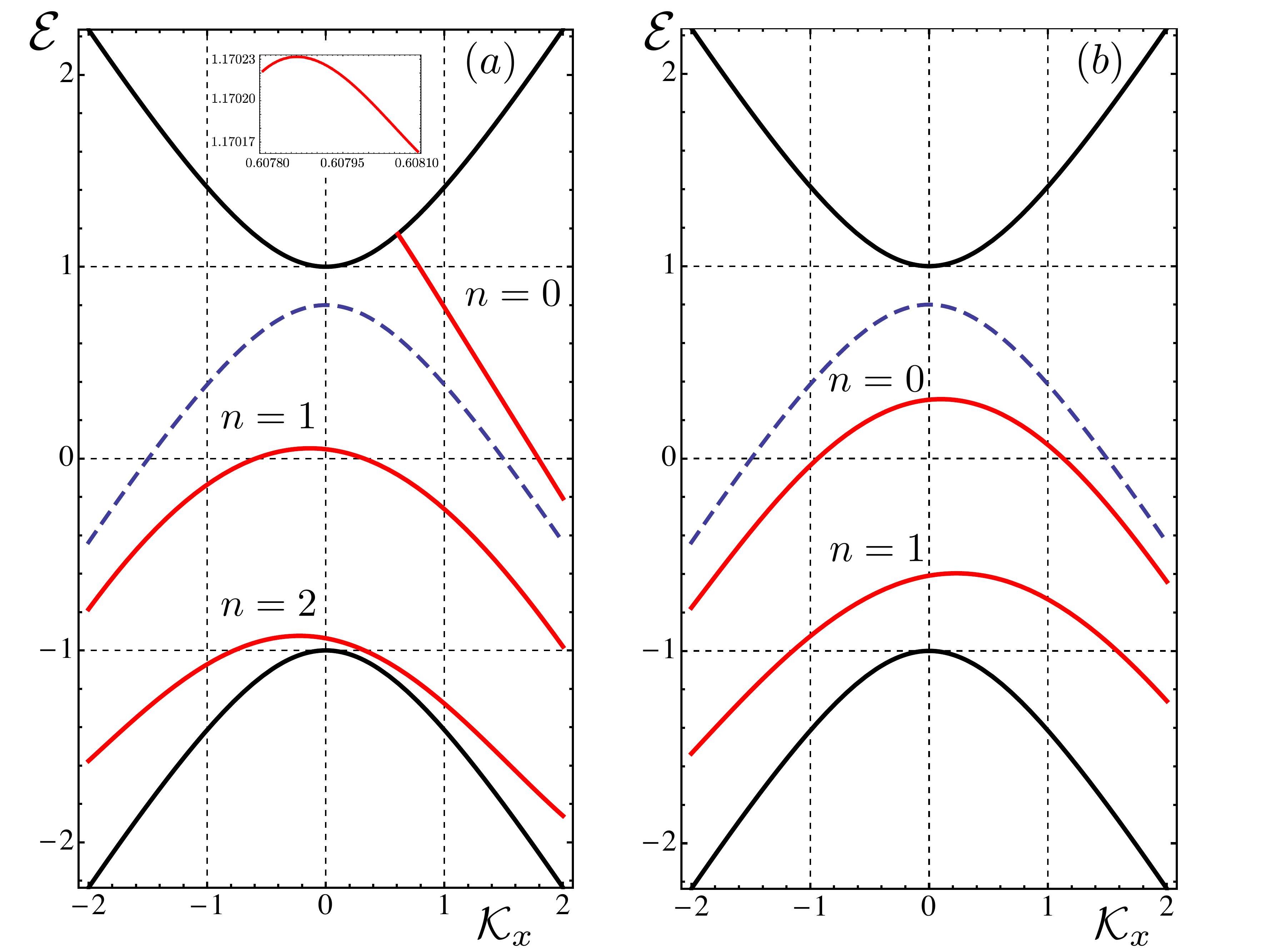}
\caption{ (Color online) Comparison of the dispersions of edge modes, shown with red, for ``topological" (a) and ``trivial" (b) boundaries. Parameters of the step are  $\tilde{U}_0=1.8$ and $\tilde{d}=2.2$.
Bulk  spectrum (black) and the spectrum in the step region (dashed) are also shown. The inset detalizes how the dispersion of the chiral mode merges with the bulk spectrum.   }
\label{figure3}
\end{figure}

\begin{figure}
\includegraphics[width=70mm]{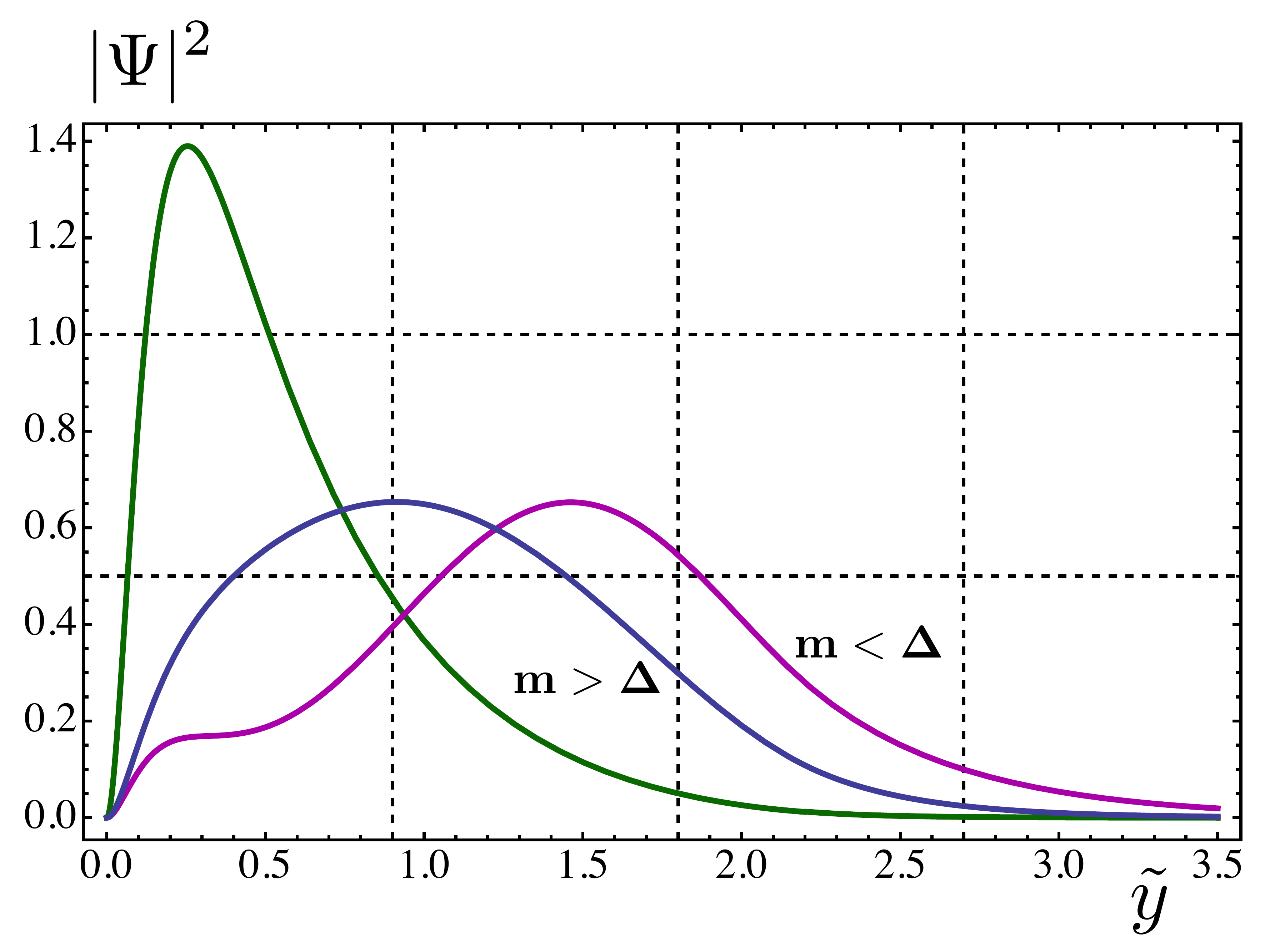}
\caption{ (Color online) Comparison of the probability density profiles for different
edge modes. For ``topological" boundary the profiles of chiral edge mode and lowest nonchiral mode are shown with green and purple, respectively. The profile for the nonchiral mode at ``trivial" boundary is shown with blue. All three profiles are calculated for
energy in the center of the gap.
  }
\label{figure4}
\end{figure}

\subsection{Chiral edge mode in the presence of  a step}
Consider a boundary with a potential step next to it depicted in  Fig. \ref{figure1}. In the domain $0<y<d$
the potential is equal to $U_0$. It creates the energy shift, so that the value $q_{\s \Delta}$
gets modified
\begin{equation}
\label{modified}
q_{\s \Delta}\rightarrow \kappa=\frac{1}{|l_{\Delta}|}\sqrt{1+(l_\Delta k_x)^2-\bigg[\frac{(E-U_0) l_\Delta}{A}\bigg]^2}.
\end{equation}
A general solution within the domain $0<y<d$ contains two growing and two decaying exponents

\begin{widetext}

\begin{align}
\label{y<d}
\Psi_{y<d} =&C_0\begin{pmatrix} -1 \\ 1\end{pmatrix} \exp\Big[i k_x x-q_0 y\Big]
+C_\Delta \begin{pmatrix} 1 \\ \frac{A(k_x +\kappa)}{m-\Delta+E-U_0}\end{pmatrix} \exp\Big[i k_x x-\kappa y\Big]+ \nonumber \\
&D_0\begin{pmatrix} 1 \\ 1\end{pmatrix}\exp\Big[i k_x x+q_0 (y-d)\Big]
+D_\Delta \begin{pmatrix} 1 \\ \frac{A(k_x -\kappa)}{m-\Delta+E-U_0}\end{pmatrix} \exp\Big[i k_x x+\kappa y\Big].
\end{align}
On the other hand, the solution for $y>d$ is still a linear combination of $\Psi_0$ and $\Psi_{\Delta}$, namely
\begin{align}
\label{y>d}
\Psi_{y>d} = C^-_0\begin{pmatrix} -1 \\ 1\end{pmatrix} \exp\Big[i k_x x-q_0 (y-d)\Big]+C^-_\Delta \begin{pmatrix} 1 \\ \frac{A(k_x +q_{\s \Delta})}{m-\Delta+E}\end{pmatrix} \exp\Big[i k_x x-q_{\s \Delta} (y-d)\Big].
\end{align}
Overall, there are $6$ unknown amplitudes in Eqs. (\ref{y<d}), (\ref{y>d}).
The $6$ boundary conditions to be satisfied is vanishing of both components of the wave function at $y=0$
and continuity of both components together with their derivatives at $y=d$.
At this point we note that the step affects the dispersion law of the edge state only for $d \gtrsim l_{\s \Delta} \gg l_0$. This observation allows for two fundamental simplifications. Firstly, the  term with amplitude $C_0$ in Eq. (\ref{y<d})
decays rapidly with $y$ from $y=0$,
so that its magnitude at the boundary $y=d$ is $\sim \exp(-d/l_0)$. Thus, this term should be taken into
account only at the boundary $y=0$. Similarly, the term with coefficient $D_0$ should be taken into account
only at $y=d$. Next, the solutions with coefficients $D_0$ and $C_0^-$  have big derivatives, $1/l_0$.
Then, the matching with the derivatives of a
slow decaying solutions, renders their amplitude  small, $\sim l_0/ l_{\Delta} \ll 1$.
Neglecting the terms $D_0$ and $C_0^-$ leaves us with the system for
$4$ unknowns with $4$ boundary conditions to satisfy. The form of this system is the following
\begin{align}
\label{system}
&-C_0+C_\Delta+D_\Delta=0, \nonumber \\
&C_0+C_\Delta \frac{A(k_x +\kappa)}{m-\Delta+E-U_0} +D_\Delta\frac{A(k_x -\kappa)}{m-\Delta+E-U_0}=0, \nonumber \\
&C_\Delta e^{-\kappa d} +D_\Delta e^{\kappa d} =C_\Delta^-, \nonumber \\
&\frac{C_\Delta (k_x +\kappa)e^{-\kappa d} +D_\Delta (k_x -\kappa)e^{\kappa d} }{m-\Delta+E-U_0}=C_\Delta^- \frac{(k_x +q_{\s \Delta})}{m-\Delta+E}.
\end{align}
The first two equations ensure that the wave function Eq. (\ref{y<d}) turns to zero at $y=0$ while the second two equations express the continuity of the wave function at $y=d$.

The consistency condition for the system Eq. (\ref{system}) yields the following
transcendental equation for the dispersion, $E(k_x)$, of the edge modes
\begin{equation}
\label{transcendenta0}
\frac{\frac{m-\Delta+E-U_0}{A(k_x-\kappa)}+1}{\frac{m-\Delta+E-U_0}{A(k_x+\kappa)}+1}e^{-2\kappa d}
=\frac{\frac{(k_x+q_{\s \Delta})(m-\Delta+E-U_0)}{(k_x-\kappa)(m-\Delta+E)}-1}
{\frac{(k_x+q_{\s \Delta})(m-\Delta+E-U_0)}{(k_x+\kappa)(m-\Delta+E)}-1}.
\end{equation}
To analyze this equation we first rewrite it in a dimensionless form

\begin{align}
\label{transcendental}
\frac{\frac{1+{\cal E}-\tilde{U}_0}{{\cal K}_x-\text{sign}(m-\Delta){\cal P}}+1}{\frac{1+{\cal E}-\tilde{U}_0}{{\cal K}_x+\text{sign}(m-\Delta){\cal P}}+1}&e^{-2 {\cal P} \tilde{d}}=
\frac{\frac{[{\cal K}_x+\text{sign}(m-\Delta){\cal Q}_{\s \Delta}](1+{\cal E}-\tilde{U}_0)}{[{\cal K}_x-\text{sign}(m-\Delta){\cal P}](1+{\cal E})}-1}
{\frac{[{\cal K}_x+\text{sign}(m-\Delta){\cal Q}_{\s \Delta}](1+{\cal E}-\tilde{U}_0)}{[{\cal K}_x+\text{sign}(m-\Delta){\cal P}](1+{\cal E})}-1},
\end{align}

\end{widetext}

where we have introduced the dimensionless energy, momentum, and the decay constant
\begin{align}
\label{dimensionless}
{\cal E}&=\frac{E}{m-\Delta},~~{\cal K}_x=\frac{A k_x}{m-\Delta}, \nonumber \\
{\cal P}&=\frac{A\kappa}{|m-\Delta|}=\sqrt{1+{\cal K}_x^2-({\cal E}-\tilde{U}_0)^2},\nonumber \\
 {\cal Q}_{\s \Delta}&=\frac{Aq_{\s \Delta}}{|m-\Delta|}=\sqrt{1+{\cal K}_x^2-{\cal E}^2},
\end{align}
while the dimensionless size and the depth of the step are defined as
\begin{equation}
\label{dimensionless1}
\tilde{U}_0=\frac{U_0}{m-\Delta},~~ \tilde{d}=\frac{|m-\Delta|d}{A}.
\end{equation}
Note that the sign of $(m-\Delta)$ appears in Eq. (\ref{transcendental}) to ensure
that the decay constant is positive for any sign of $(m-\Delta)$.

The dispersion law Eq. (\ref{chiral}) for the chiral edge state follows from Eq. (\ref{transcendental}) in the limit
$\tilde{d}\rightarrow 0$.  Indeed, in dimensionless units,  Eq. (\ref{chiral}) reads
${\cal E}=-{\cal K}_x$. This suggests that ${\cal Q}_{\s \Delta}=1$. For $(m-\Delta)<0$, the ratio $(1+{\cal E})/({\cal K}_x+\text{sign}(m-\Delta){\cal Q}_{\s \Delta})$ is equal to
$-1$, the fractions in the left-hand side and in the right-hand side are equal to each other, so that Eq. (\ref{transcendental}) is satisfied.  It is even easier to see that in the limit $\tilde{d}\rightarrow \infty$ Eq. (\ref{transcendental}) yields the dispersion law
${\cal E}=-{\cal K}_x+{\tilde U}_0$. In this limit the denominator in the left-hand side
turns to zero for negative $(m-\Delta)$.

 For general parameters of the step the dispersion law of the chiral mode
is illustrated in Figs. \ref{figure2}, \ref{figure3}.
 Naturally, presence of the step does not violate the fact that
 the chiral  mode exists  only for negative $(m-\Delta)$.
 For a ``weak" step the edge mode is present for both positive and negative
momenta, while for a ``strong" step only at positive momenta. Although it is not a rigorous
statement, the dispersion is linear with very high accuracy. Numerically, the relative change of the slope with ${\cal K}_x$ is $\approx 10^{-3}$.

 Figs. \ref{figure2}, \ref{figure3} also suggest that
the dispersion  of a chiral edge mode has an endpoint. This is also the consequence of a finite
accuracy of the numerical procedure. The true behavior of the slope, as the edge mode
merges with continuum at certain point ${\cal K}_x={\cal K}_x^c$,
${\cal E}={\cal E}^c=\left[1+({\cal K}_x^c)^2\right]^{1/2}$
is
$\big[\frac{\partial {\cal E}}{\partial {\cal K}_x}- \frac{{\cal K}_x^c}{{\cal E}^c}\big]\propto \left({\cal K}_x-{\cal K}_x^c\right)$. To see this, one can view the transcendental equation Eq. (\ref{transcendental}) as a relation between the variables ${\cal K}_x$ and ${\cal Q}_{\s \Delta}$. Since it contains the terms linear in ${\cal K}_x$ and ${\cal Q}_{\s \Delta}$,
its variation yields $\delta {\cal K}_x=\eta \delta {\cal Q}_{\s \Delta}$, where $\eta$ is
some constant. On the other hand, from definition of ${\cal Q}_{\s \Delta}$ it follows that
${\cal Q}_{\s \Delta}\delta {\cal Q}_{\s \Delta}={\cal K}_x \delta{\cal K}_x-{\cal E}\delta{\cal E}$. Thus, one has
\begin{equation}
\label{relation}
\frac{{\cal Q}_{\s \Delta}}{{\cal E}}=\eta\Big(\frac{{\cal K}_x}{{\cal E}}-\frac{\partial {\cal E}}{\partial {\cal K}_x}\Big).
\end{equation}
As the dispersion law approaches the continuum, the variable   ${\cal Q}_{\s \Delta}$ turns to zero. Then it follows from Eq. (\ref{relation}) that the velocity of the edge mode approaches  $\frac{{\cal K}_x^c}{{\cal E}^c}$, which is the velocity of the bulk mode. Numerically, the merging of the chiral edge
mode dispersion with the bulk spectrum is illustrated in Fig. \ref{figure3}, inset. It is seen that the change of
sign of the slope takes place within a very narrow domain of momenta $\sim 10^{-4}$.

\subsection{Nonchiral edge modes}

Our main finding in the present paper is that the transcendental equation Eq. (\ref{transcendental})
captures, along with the chiral mode,  a set of nonchiral edge modes. For these modes the decay rate,
$\kappa$, within the step and, thus, the dimensionless ${\cal P}$ are purely imaginary. For such ${\cal P}$ it is convenient to cast Eq. (\ref{transcendental}) in the form
\begin{equation}
\label{transcendental2}
\vert {\cal P}\vert {\tilde d}+ \Phi_1+\Phi_2 =\pi n,
\end{equation}
where $n=0, 1, 2, ....$ is integer and the phases $\Phi_1$, $\Phi_2$ are defined as
\begin{align}
\Phi_1&= \arctan\Bigg(\frac{\text{sign}(m-\Delta)\vert {\cal P}\vert}{1+{\cal E}-\tilde{U}_0+{\cal K}_x}\Bigg),\nonumber \\
\Phi_2&=  \arctan\Bigg(\frac{\text{sign}(m-\Delta)\vert {\cal P}\vert}{\frac{1+{\cal E}-\tilde{U}_0}{1+{\cal E}}({\cal K}_x+\text{sign}(m-\Delta){\cal Q}_{\s \Delta})-{\cal K}_x} \Bigg)   .
\end{align}
The meaning of $\vert {\cal P}\vert \tilde{d}$ is the phase accumulated by the components of the pseudospinor on the interval $0<y<d$,  where they oscillate. The meaning of $\Phi_1$ and $\Phi_2$
is the phase shift at the boundary $y=0$ and $y=d$, respectively.

Both phase shifts depend on sign of $(m-\Delta)$. Thus, the dispersion laws of nonchiral  modes
``know" whether or not the band structure is inverted. These dispersion laws,
 obtained from Eq. (\ref{transcendental2}) are shown in Fig. \ref{figure3}  for a given step
 and with opposite signs of $(m-\Delta)$. It is seen that for $(m-\Delta)>0$ the nonchiral branches
 lie {\em deeper in the gap} than nonchiral branches for negative $(m-\Delta)$. The sign of $(m-\Delta)$
 also determines the classification of the branches. For $(m-\Delta)>0$ the values of $n$ start from
$n=0$, while $(m-\Delta)>0$ they start from $n=1$. Qualitatively, this suggests that a chiral mode
``complicates" the formation of nonchiral modes.
Different dispersions for positive and negative $(m-\Delta)$ implies that the behavior of $\vert \Psi (y) \vert^2$ is different. This is illustrated in Fig. \ref{figure4}. We see that nonchiral mode for
$(m-\Delta)<0$ is significantly more extended than for $(m-\Delta)>0$.

\begin{figure}
\includegraphics[width=84mm]{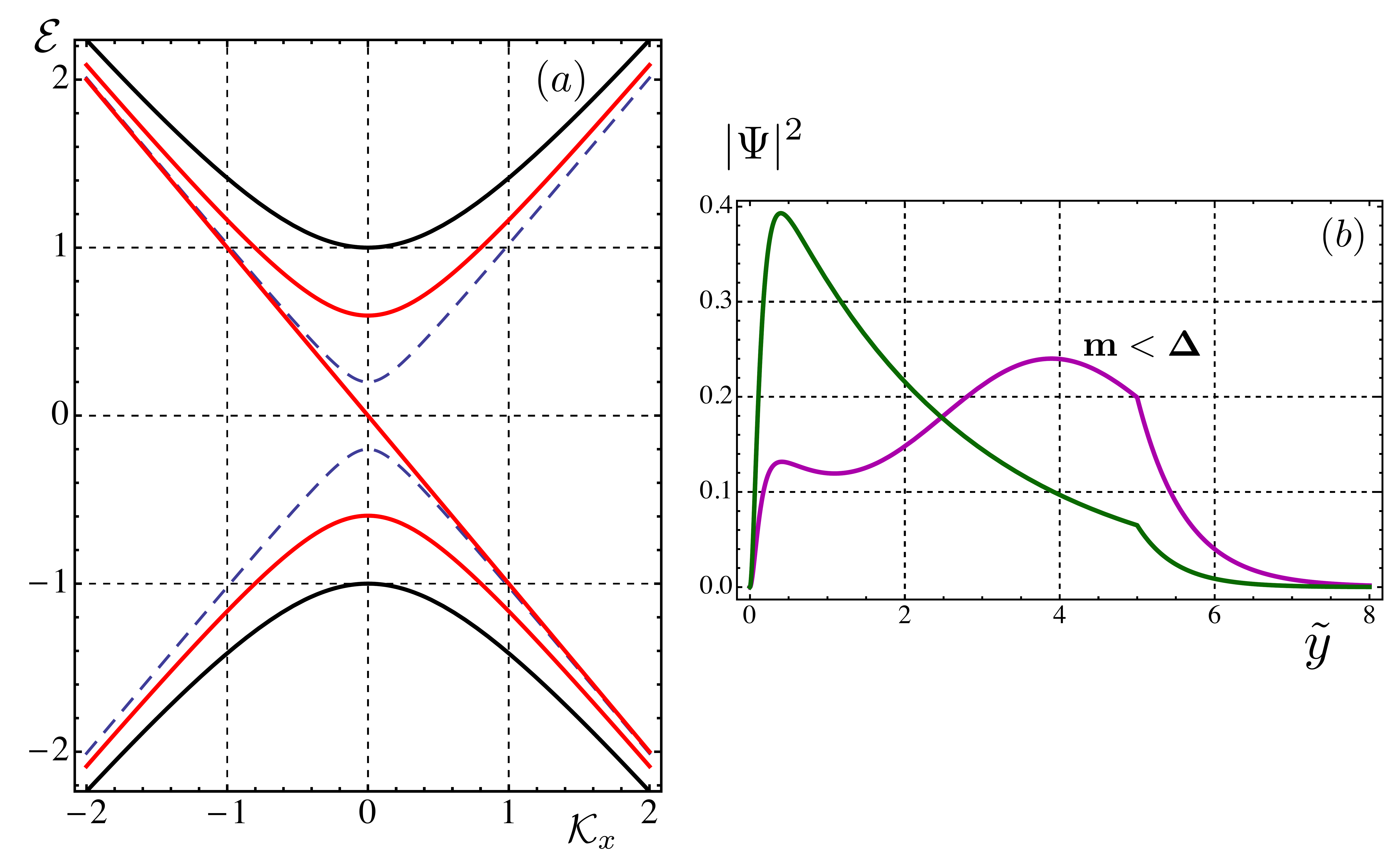}
\caption{(Color online) Dispersions (a) and the density profiles (b) of a chiral and
 nonchiral modes are shown for the extended  confinement Fig. 1b. The magnitude of the gap
 reduction near the edge is $\delta=0.2\Delta$, while the dimensionless width is $\tilde{d}=5$. Density profile of both modes is calculated for dimensionless energy ${\cal E}=0.6$.
 Bulk  spectrum (black) and the spectrum in the step region (dashed) are also shown.}
\label{figure5}
\end{figure}

It is instructive to compare the above results for the step potential with dispersion of nonchiral
modes emerging from a jump of the gap magnitude next to the boundary in the domain $0<y<d$, see Fig. \ref{figure1}b.
Modifications of Eq. (\ref{transcendental}) to this case are straightforward. Firstly, the decay constant ${\cal P}$
should be redefined
\begin{equation}
{\cal P}\rightarrow \sqrt{\tilde{\delta}^2+{\cal K}_x^2-{\cal E}^2},
\end{equation}
where $\tilde{\delta}=(m-\delta)/(m-\Delta)$ is the relative reduction of the gap in the domain $0<y<d$.
The second modification is the replacement of the combination $1-\tilde{U}_{\s 0}$ in Eq.  (\ref{transcendental})
by $\tilde{\delta}$. The solutions of Eq.  (\ref{transcendental}) for a particular set of parameters are shown in
Fig. \ref{figure5}a. Naturally,  nonchiral modes are symmetric with respect to $E=0$. Unlike the case of potential
step, they never reach the midgap. With regard to the density profile, Fig. \ref{figure5}b, the nonchiral mode is
repelled from the boundary even further than in the case of potential step.

\section{``Topological" shift of the dispersion of the localized bulk mode}

Suppose that instead of a step there is a potential, $U(y)$,  well separated from the boundary
by a distance $D \gg l_{\s \Delta}$. In the limit $D\rightarrow \infty$ the  dispersion, $E_{\s 0}(k_x)$, of a nonchiral mode, corresponding to the bound state in $U(y)$,
does not depend on whether or not the underlying band structure is inverted. For a finite $D$
the presence of the boundary will manifest itself as correction $\delta E_{\s 0}(k_x)$
to the dispersion law. From the above consideration of the step confinement, it is apparent
that this correction has a ``topological" character: it shifts $E_{\s 0}(k_x)$ towards
the center of the gap for $(m-\Delta)>0$ and away from the midgap for $(m-\Delta)<0$. In the limit
of large $D$ this correction can be found perturbatively in parameter $l_{\s \Delta}/D$.

Denote with $\psi_e(y)$ and $\psi_h(y)$ the components of pseudospinor describing the wave function
of a nonchiral mode

\begin{widetext}

\begin{equation}
\label{withwall}
\Big(\hat{h}^\ast(-{\bf \hat{k}})+U(y)\Big)\Psi=
\Big(\hat{h}^\ast(-{\bf \hat{k}})+U(y)\Big)\begin{pmatrix} \psi_e \\ \psi_h \end{pmatrix}
=E(k_x) \begin{pmatrix} \psi_e \\ \psi_h \end{pmatrix}.
\end{equation}
For $l_0 \ll l_{\s \Delta}$ presence of the boundary is taken into account by imposing a boundary  condition
\begin{equation}
\label{condition}
\Psi(0)=\begin{pmatrix} 1 \\ -1 \end{pmatrix}.
\end{equation}
We emphasize that, as in the case of a step, the meaning of $y=0$ in this condition is, in fact,
$l_0 \ll y \ll l_{\s \Delta}$.
 Denote now with $\psi_e^{(0)}(y)$,  $\psi_h^{(0)}(y)$
the component of pseudospinor for the case when the boundary is absent (moved to $y=-\infty$).
One has
 \begin{equation}
 \label{withoutwall}
\hspace{-4mm}\Big(\hat{h}^\ast(-{\bf \hat{k}})+U(y)\Big)\Psi^{(0)}=
\Big(\hat{h}^\ast(-{\bf \hat{k}})+U(y)\Big)\hspace{-1mm}\begin{pmatrix} \psi_e^{(0)} \\ \psi_h^{(0)} \end{pmatrix}
=E_{\s 0}(k_x)\hspace{-1mm} \begin{pmatrix} \psi_e^{(0)} \\ \psi_h^{(0)} \end{pmatrix}.
\end{equation}
As a next step, we multiply Eq. (\ref{withwall}) by $\Psi^{(0)}$ from the left
and Eq. (\ref{withoutwall}) by $\Psi$ from the left and subtract them from each other.
This yields
\begin{equation}
\label{subtracted}
A\frac{d( \psi_e^{(0)} \psi_h- \psi_e \psi_h^{(0)})}{dy}=\delta E_{\s 0}(k_x) \Big( \psi_e^{(0)}(y) \psi_e(y)+ \psi_h (y)\psi_h^{(0)}(y)\Big).
\end{equation}
Upon integrating Eq. (\ref{subtracted}) from $y=0$ to $\infty$, we find the analytical expression
for $\delta E_{\s 0}(k_x)$
\begin{equation}
\label{integrated}
\delta E_{\s 0}(k_x) =-A\frac{ \psi_e^{(0)}(0) \psi_h(0)- \psi_e(0) \psi_h^{(0)}(0)}
{\int\limits_{0}^{\infty} dy \Big( \psi_e^{(0)}(y) \psi_e(y)+ \psi_h (y)\psi_h^{(0)}(y)\Big)}.
\end{equation}
The difference between the boundary values $\psi_e(0)$ and $\psi_e^{(0)}(0)$ as well as
$\psi_h(0)$ and $\psi_h^{(0)}(0)$ is that the exact wave functions satisfy the
boundary condition Eq. (\ref{condition}). The boundary leads to the admixture to $\Psi^{(0)}$
of the ``short-range" solution decaying into the bulk as $\exp(-q_0y)$ and of the ``reflected"
solution decaying as $\exp(-q_{\s \Delta}y)$. The corresponding amplitudes, $C_0$ and $C_{\s \Delta}$, are found from the boundary condition
\begin{equation}
\label{boundary2}
C_0\begin{pmatrix} -1 \\ 1\end{pmatrix}
+C_\Delta \begin{pmatrix} 1 \\ \frac{A(k_x +q_{\s \Delta})}{m-\Delta+E_{\s 0}}\end{pmatrix} + \begin{pmatrix} \psi_e^{(0)}(0) \\ \psi_h^{(0)}(0) \end{pmatrix}=0,
\end{equation}
which yields
\begin{equation}
\hspace{-4mm}C_0=\frac{-\frac{A(k_x +q_{\s \Delta})}{m-\Delta+E_{\s 0}}\psi_e^{(0)}(0)+\psi_h^{(0)}(0)}{1+\frac{A(k_x +q_{\s \Delta})}{m-\Delta+E_{\s 0}}},~ C_{\Delta}=-\frac{\psi_e^{(0)}(0)+\psi_h^{(0)}(0)}{1+\frac{A(k_x +q_{\s \Delta})}{m-\Delta+E_{\s 0}}}.
\end{equation}
At distance $y\gg l_0$ from the boundary the short-range solution vanishes. Thus, the
differences $\psi_h(0)-\psi_h^{(0)}(0)$, and  $\psi_e(0)-\psi_e^{(0)}(0)$
are determined only by the reflected solution
\begin{equation}
\label{Psi2}
\begin{pmatrix} \psi_e(0) \\ \psi_h(0)\end{pmatrix}=
C_\Delta \begin{pmatrix} 1 \\ \frac{A(k_x +q_{\s \Delta})}{m-\Delta+E_{\s 0}}\end{pmatrix} + \begin{pmatrix} \psi_e^{(0)}(0) \\ \psi_h^{(0)}(0) \end{pmatrix}.
\end{equation}
Substituting Eq. (\ref{Psi2}) into Eq. (\ref{integrated}), we express the correction
$\delta E_{\s 0}(k_x)$ via the components of the bare pseudospinor
\begin{equation}
\label{final}
\delta E_{\s 0}(k_x) =-A\frac{ \psi_e^{(0)}(0) \psi_h^{(0)}(0)\Big(1-\frac{k_x +q_{\s \Delta}}{k_x -q_{\s \Delta}}\Big)\Big(\frac{1+\frac{A(k_x-q_{\s \Delta})}{m-\Delta+E_{\s 0}}}{1+\frac{A(k_x +q_{\s \Delta})}{m-\Delta+E_{\s 0}}}\Big)}
{\int\limits_{-\infty}^{\infty} dy \Big[ (\psi_e^{(0)}(y))^2 + (\psi_h^{(0)}(y))^2\Big]},
\end{equation}
where we took into account that $\psi_h^{(0)}(0)/\psi_e^{(0)}(0)=A(k_x -q_{\s \Delta})/(m-\Delta+E_{\s 0})$.

We see that the correction is proportional  to the product of the bare amplitudes, and thus to $\exp(-2q_{\s \Delta}D)$, which is the probability
to find an electron at the edge. The result Eq. (\ref{final}) applies when this probability is small.
For this reason we replaced $\psi_e(0)$, $\psi_h(0)$ in the denominator by $\psi_e^{(0)}(0)$, $\psi_h^{(0)}(0)$ and extended the low limit of integration to $-\infty$.
To analyze the dependence of the correction on the bare spectrum, $E_{\s 0}(k_x)$, it is instructive
to recast the last bracket into the form

\begin{align}
\frac{m-\Delta+E_{\s 0}+A(k_x-q_{\s \Delta})}{m-\Delta+E_{\s 0}+A(k_x +q_{\s \Delta})}=\frac{\Big[m-\Delta+E_{\s 0}+Ak_x-\sqrt{(m-\Delta)^2+A^2k_x^2-E_{\s 0}^2}\Big]^2}{2(E_{\s 0}+Ak_x)(m-\Delta+E_{\s 0})}.
\end{align}

\end{widetext}

The above expression illustrates the topological origin of the shift of a nonchiral mode,
$E_{\s 0}(k_x)$. Indeed, the correction Eq. (\ref{final}) contains a pole corresponding to the
dispersion law of the chiral edge mode.
This confirms our earlier observation that presence of this mode complicates the formation of nonchiral modes. The shift Eq. (\ref{final}) tends to reduce the binding energy. Another feature
that points at the topological origin of the correction is that it depends on the sign of $k_x$.
This is in contrast to non-perturbed behavior $E_{\s 0}(k_x)$ which is an even function of $k_x$. As $k_x$ increases, the  parameter $q_{\s \Delta}$, which is the characteristics of proximity of
$E_{\s 0}(k_x)$ to the continuous spectrum, becomes much smaller than $k_x$. Then the second
bracket in Eq. (\ref{final}) is close to $1$, while the first bracket falls off as $1/k_x$.

The result Eq. (\ref{final}) strongly simplifies for small $k_x$. Then we have
\begin{align}
\label{final1}
\delta E_{\s 0}(k_x) =&-A\frac{2 \psi_e^{(0)}(0) \psi_h^{(0)}(0)}
{\int\limits_{-\infty}^{\infty} dy \Big[ (\psi_e^{(0)}(y))^2 + (\psi_h^{(0)}(y))^2\Big]}
\nonumber\\
&\times \Bigg[\frac{(m-\Delta)-\sqrt{(m-\Delta)^2-E_{\s 0}^2}}{E_{\s 0}}\Bigg].
\end{align}
It is the factor in the square brackets that carries  information on whether the or not
the boundary supports the chiral mode. Indeed, if the level, $E_{\s 0}$, in the potential $U(y)$
is close to midgap, then this factor diverges for $(m-\Delta)<0$, while it turns to zero for $(m-\Delta)>0$. This is because, for $(m-\Delta)<0$, there is a level $E=0$ at the boundary from which the level $E_{\s 0}$ is repelled. When this level is absent, the behavior of the shift $\delta E_{\s 0}\propto E_{\s 0}$ is natural. For $E_{\s 0}\rightarrow 0$ there are equal probabilities to be shifted up or down.

\section{Concluding remarks}

(i) Presence or absence of chiral modes in QAH effect is decided by the relative sign of
$(m-\Delta)$ and parameter $B$ in the Hamiltonian ${\hat h}({\bf k})$, although the parameter
$B$ itself does not enter into the dispersion law of the chiral mode. The situation with nonchiral
modes is analogous.  While it does not enter into their dispersion relations, these relations depend
on whether $(m-\Delta)$ and $B$ have the same sign or opposite signs. Moreover, similarly to chiral
mode, nonchiral modes will not exist without the term $Bk^2$ on the diagonal of the matrix ${\hat h}$.
This is because without the short-range solution $\propto \exp(-q_0y)$ in Eq. (\ref{y<d}) the hard-wall
boundary conditions cannot be satisfied.

(ii) Within the standard picture of the QAH transition\cite{2008} it takes place as
the gap closes and two chiral modes at the opposite edges merge. In this regard, our
main finding is that these modes can ``communicate" with each other via nonchiral edge modes
which are less localized. In other words, nonchiral modes emerging as a result of the
extended  confinement smear the QAH transition.

(iii) Our other finding is that, while nonchiral modes are present for both signs of $(m-\Delta)$,
their formation is much less likely for  $(m-\Delta)<0$. This can be interpreted as follows.
The pseudospinor corresponding to  nonchiral mode should be orthogonal to the chiral mode,
if it is present. Thus the formation of nonchiral mode is impeded for ``topological" sign of
$(m-\Delta)$.

(iv) In Ref. \onlinecite{VolkovBoundary} it was assumed that the boundary of the system is planar,
and the generalized, compared to hard wall, version of the boundary conditions was employed.
It was demonstrated that variation of parameters in the boundary condition can lead to disappearance
of the chiral mode form the gap, but nonchiral modes do not emerge upon this variation.

(v) It is straightforward to generalize our results for rectangular step to arbitrary profile of
the step. Essentially, the decay constant $\kappa$ defined by Eq. (\ref{modified}) becomes the
function of coordinate. Qualitative
conclusions do not change.

(vi) With regard to the effect of nonchiral modes on quantization of the components of the resistivity tensor
in realistic Hall-bar device, quantitative analysis in Ref. \onlinecite{EdgeTransportQAH} based on the
Landauer-B{\"u}ttiker approach indicates that, due to backscattering of nonchiral modes, the quantization of the diagonal component
is violated, while the non-diagonal component remains nearly quantized.

\section{Acknowledgements}

We are grateful to D. A. Pesin for a number of illuminating discussions. We are grateful to
Jing Wang (Stanford University) for introducing  the topic  of  QAH effect to us.  This work was supported by NSF through MRSEC DMR-1121252.


\end{document}